\RequirePackage{lineno}
\documentclass[aps,preprint,floatfix]{revtex4}
\usepackage{amsmath,units,xspace,amssymb}
\usepackage[pdftex]{graphicx}
\usepackage{booktabs}
\usepackage{braket}
\usepackage{xcolor}
\usepackage[normalem]{ulem}
\usepackage[labelformat=empty]{caption}
\begin{document}


\title{Anatomy of cage formation in a 2D glass-forming liquid}
\author{Bo Li,$^{1}$ Kai Lou,$^{1}$ Walter Kob$^{2,3,\star}$ and Steve Granick$^{1}$}
\affiliation{1, Center for Soft and Living Matter, Institute of Basic Science, Ulsan, 44919, South Korea \\ 2, Laboratoire Charles Coulomb, University of Montpellier, CNRS, 34095 Montpellier, France\\ 3, Institut Universitaire de France}
\date{\today}

\begin{abstract} 
The solidity of glassy materials is believed to be
due to the cage formed around each particle by its
neighbors, but in reality the details of cage-formation remain
elusive~\cite{2011Glassy_Binder,2002_Weeks,Sastry_2009,Pastore_2017}. This
cage starts to be formed at the onset temperature/density at which the
normal liquid begins to show the first signs of glassy dynamics. To study
cage-formation we use here focused lasers to produce a local perturbation
of the structure on the particle level in 2D colloidal suspensions
and monitor by means of video microscopy the system's non-linear
dynamic response. All observables we probed show a response which is
non-monotonic as a function of the packing fraction, peaking at the onset
density. Video microscopic images reveal that this maximum response is
due to the buildup of domains with cooperative dynamics that become
increasingly rigid and start to dominate the particle dynamics. This
proof-of-concept from microrheological deformation  demonstrates that
in this glass-forming liquid cage formation is directly related to
the merging of these domains, thus  elucidating the first step in
glass-formation~\cite{2011Glassy_Binder,2009_Cavagna}.

\end{abstract}

\maketitle

\noindent
Corresponding authors email:
$^\star$ walter.kob@umontpellier.fr
\\[20mm]

Despite broad agreement that glasses are important intellectually and
technologically, we have at present little insight about the nature of
the particle dynamics at the onset point, i.e.~where the dynamics crosses
over from normal liquid dynamics to a glassy one, i.e.~time correlation 
functions are stretched, the dynamics starts to become heterogeneous, etc.~\cite{2011Glassy_Binder}.
Usually this point is
detected from the appearance of a plateau in time correlation functions
such as the mean squared displacement (MSD) or the intermediate scattering
function, i.e. the dynamics starts to show a two-step relaxation
\cite{2011Glassy_Binder,1993_vanmegen,1990_barrat,1995PRE_Kob}. However,
being ensemble-averaged, these quantities do not give access to
information about local particle dynamics and hence very little is
known about cage-formation on the level of the particles. Direct
imaging experiments do allow to access the local particle dynamics
\cite{1996ARPC_Murray,2000Science_Weeks,2000Science_Kegel,2018CP_Li}
but were previously not coupled to methods capable of studying the
inelastic response of the structure to local perturbation, i.e. to probe
the local dynamics in a precise manner (see Ref.~\cite{2013JCP_Anderson}
for experiments on the local elastic response). For theory the onset point
is a challenging state since it corresponds to the boundary at which the
fast liquid dynamics of a normal fluid meets the phonon dynamics relevant
in the short time regime of the glassy relaxation. The competition
between these two kinds of dynamics makes an analytical calculation most
difficult because of the presence of both, fast and slow, processes. These
considerations may explain why, to the best of our knowledge, the study
of this dynamic regime on the level of individual particles was not
previously presented. In the present work we thus use focused laser
beams to probe the relaxation dynamics by imaging the response to local
deformations of individual particles and hence elucidate the surprisingly
complex formation of the cage on the level of the particles.

Colloidal systems display many properties of atomic glass-formers with
the advantage that optical imaging can track directly the trajectories
of individual particles \cite{1996ARPC_Murray,2012RPP_Hunter}. Focused
laser beams allow to manipulate the colloidal system on the level
of the particles \cite{2015NP_Nagamanasa} and hence to measure
local responses. This has been exploited to probe the elastic
behavior of colloidal  glasses \cite{2013JCP_Anderson} and crystals
\cite{2017PNAS_Buttinoni, 2018PRL_Cash, 2018PNAS_Lavergne,2019_Lozano}
and here we extend their use to study the micro-rheological properties
of the system with deformations beyond the linear regime. We consider a
two-dimensional binary colloidal system of spherical particles (PMMA,
polymethylmethacrylate, in water) that are confined between two glass
plates separated by a distance of 3.37~$\mu$m. The concentration and
size ratio between the small (s) and large (l) particles are 0.55:0.45
and 2.08~$\mu$m:2.91~$\mu$m, respectively, making the system a good
glass-former that is not prone to crystallisation, see Extended Data
Fig.~2{\bf a)}. We study the behavior of the liquid as a function of
the area packing fraction $\phi=\frac{1}{4} \pi (\sigma_\textrm{s}^{2}
\rho_\textrm{s}+\sigma_\textrm{l}^{2} \rho_\textrm{l})$, where
$\sigma_{\rm s}$ [$\sigma_{\rm l}$] and $\rho_{\rm s}$ [$\sigma_{\rm
l}$] are the size and number density of the small [large] partices,
respectively. Experimental details are described in Methods where we
also show that the onset packing fraction $\phi_\textrm{onset}$ is around
0.60 and the critical density of mode-coupling theory~\cite{2008_Gotze},
$\phi_\textrm{MCT}$, is above 0.75 (Extended Data Fig.~2). We determine
the reponse of the system in the range $0.45 \leq \phi \leq 0.83$ thus
well below the glass transition which occurs around 0.90 (see Methods). From previous work it is known that 2D systems such
as ours capture the essence of the glass problem because the mechanism
leading to caging does not depend strongly on the dimensionality of the
system~\cite{2011Glassy_Binder,2012RPP_Hunter,2017PNAS_Vivek,2019PNAS_Flenner}.

\begin{figure}[!h]
\centering
\includegraphics[width=1.0\columnwidth]{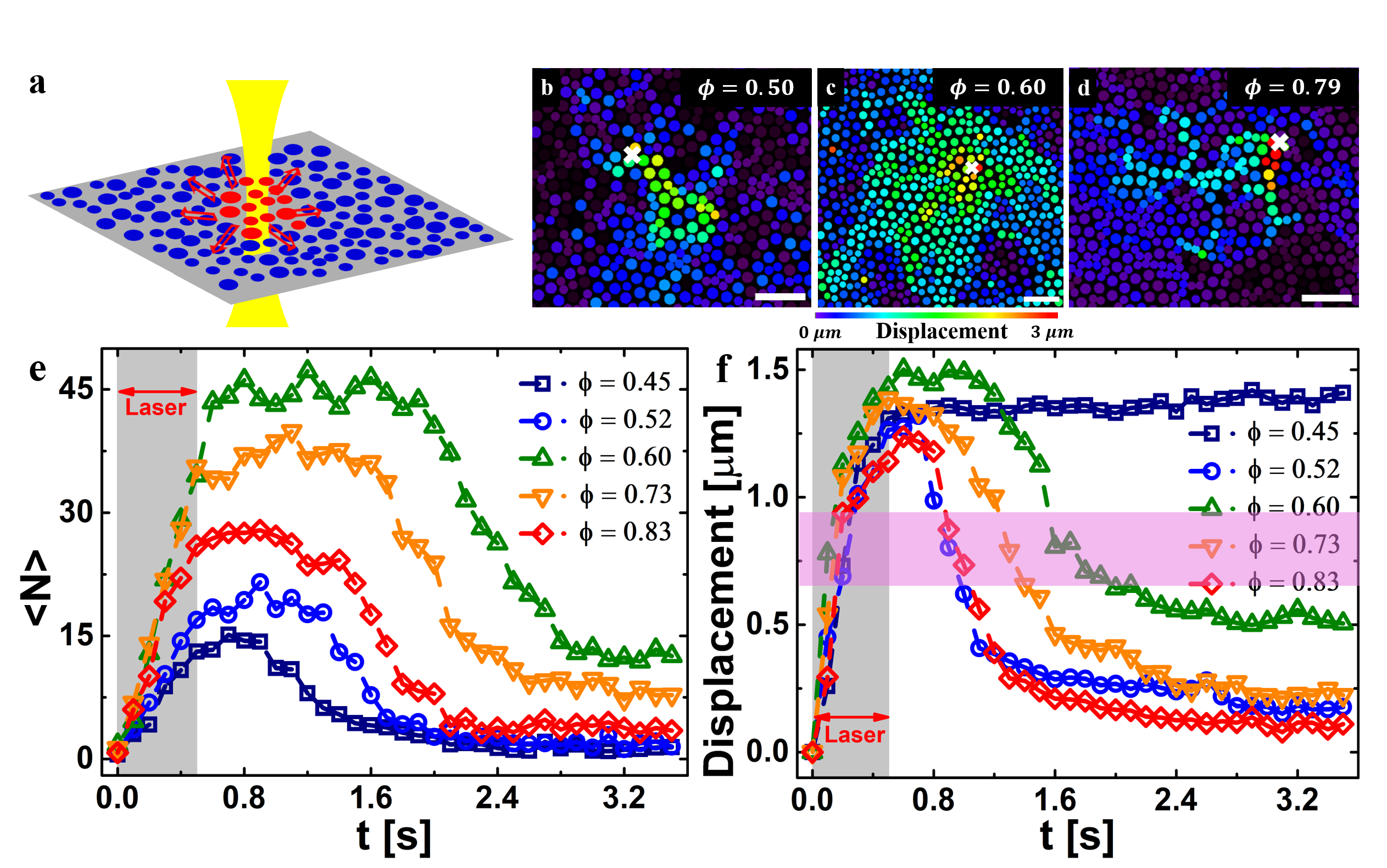}
\caption{\textbf{Figure 1 $|$ The experimental system and responses to
local perturbations.} \textbf{a)} Schematic view showing that a laser
pulse (yellow) causes local displacements (red). \textbf{b)}-\textbf{d)}
Typical particle displacements 5~s after the laser pulse has been
turned off for 3 packing fractions specified in the panels. Scale
bar is 10~$\mu$m. The color map is specified below panel \textbf{c)}.
\textbf{e)} Average number of particles participating in an excitation
for different packing fractions, plotted against time. \textbf{f)} Average
displacement, $\delta$, of the particles in the excitation. Note that the
maximum value of $\langle N \rangle$ and $\delta$ is non-monotonic as a
function of $\phi$. The shaded pink area indicates the size of the cage
at $\phi=0.60$. In panels {\bf b)} - {\bf f)} the laser power is $99$~mW.}
\label{fig:1}
\end{figure}

Figure~\ref{fig:1}{\bf a)} summarises the experiment schematically. To
create local perturbations, we use pulsed laser beams (repetition
rate 80 MHz) of duration 0.5~s focused to a spot of size 2.0~$\mu$m,
i.e.~comparable to the size of the individual colloids. The strong
electric field gradient generates a dielectrophoresis force, whose
magnitude is proportional to the laser intensity $A$, acting on the
particle in the direction of the gradient and thus induces a rapid motion
of the colloids hit by the beam (see Ref.~\cite{2017PNAS_Buttinoni},
Methods and Video~1 for details) which in turn collide with their
neighbors, hence setting up a local motion that in the following will
be called ``excitation''. In the Methods section we demonstrate that
the zone influenced directly by the dielectrophoresis force is small ($<
1.2~\sigma_{s}$) and independent of $A$ and that the pulse duration and
exact position the beam hits the particle does not alter the qualitative
properties of the excitation pattern. Figures~\ref{fig:1}{\bf b)-d)}
show the displacement of the particles 5~s after the laser pulse,
i.e.~when the excitation has stopped. (Excitations last less than
3.5~s, Fig.~\ref{fig:1}{\bf e)}, i.e.~one order of magnitude shorter
than the typical $\alpha$-relaxation time of the system in the density
regime we probe, see Methods.) At low packing fraction, $\phi=0.50$
panel {\bf b)}, only few particles are displaced and most relax back to
their initial position (see Video~2). If $\phi$ is increased to 0.60,
the number of displaced particles has increased significantly, panel
{\bf c)} and Video~3, while this number decreases again if $\phi$ is
0.70, Video~4. In the Methods we show that an excitation is {\it not}
a random  structure, since repeated perturbations at the same spot give
rise to excitations with similar structures if the laser intensity is
less than $\approx$44~mW (Extended Data Figs.~5{\bf a)-c)}), although
even below this threshold the perturbation is non-linear. These results
show that the formation of the cage is a surprisingly non-local process
involving particles that are not only in the first nearest neighbor shell.

To probe the properties of the excitations, we analyse for each value
of $\phi$ dozens of displacements within circular zones of radius
7~$\sigma_{\rm s}$ - the precise size is arbitrary, but was selected to
be larger than the particle diameter itself, yet small enough to avoid a
significant contribution from particles unaffected by the excitation. The
impact of the perturbation is quantified by means of two quantities,
the number of mobile particles $N$ (those particles that move more
than 0.3~$\sigma_\textrm{s}$ after the laser pulse was turned on)
and $\delta$, the average displacement of these particles. For small
$\phi$, $\langle N\rangle$ remains small and $\delta$ quickly reaches
a high value and then remains large, showing the viscous response
to the perturbation, Fig.~\ref{fig:1}{\bf e)} and {\bf f)}. (Here
$\langle . \rangle$ is the average over the excitations.) At $\phi$
= 0.52 one finds that both quantities grow but then return to lesser
values - there is some elasticity. At $\phi$ = 0.60, the peaks are even
higher and broader with $\langle N \rangle$ reaching 45 particles,
before the signals decay to a finite plateau. At even higher $\phi$,
the peak height in $\langle N \rangle$ and $\delta$ the displacement,
as well as their values at long times are {\it smaller} than the ones
for $\phi=0.60$. This non-monotonic behavior of the two observables is
striking since sample-averaged quantities such as the viscosity or the
relaxation time increase monotonically for all packing fractions. Also
included in the graph is the size of the cage as obtained from the height
of the plateau of the MSD at $\phi=0.6$, see Extended Data Fig.~2{\bf
c)}. One recognizes that at long times the average displacement of the
particles is somewhat smaller than the size of this cage, in agreement
with the results presented below on the van Hove function.

\begin{figure}[!h]
\centering
\includegraphics[width=1.0\columnwidth]{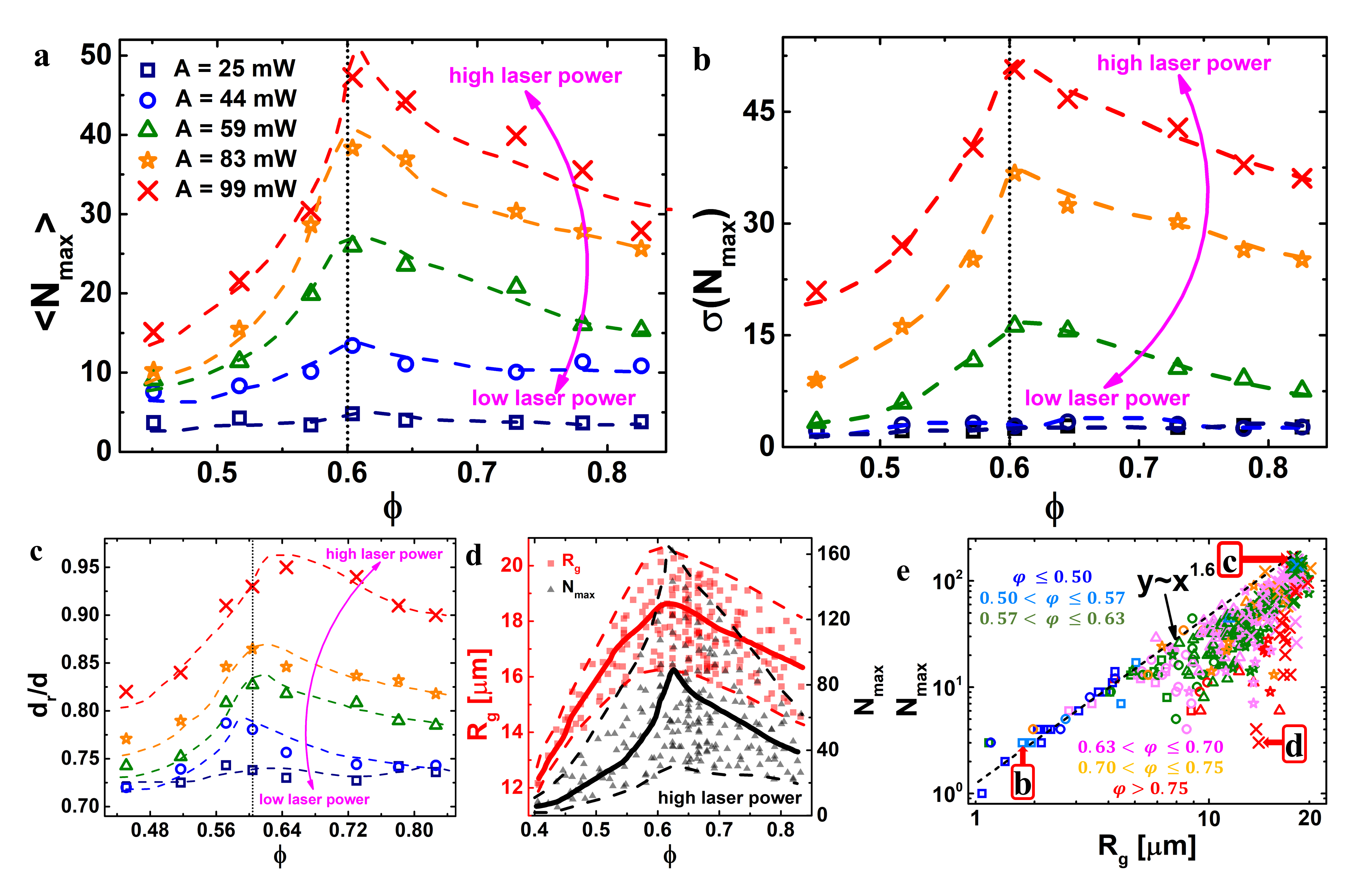}
\caption{\textbf{Figure 2 $|$ Properties of excitations are non-monotonic
in packing density.} The mean maximum number of particles participating in
an excitation, $\braket{N_{\rm max}}$, \textbf{a)}, and its variance,
\textbf{b)}, plotted against $\phi$ at different laser powers. \textbf{c)}
The radial component $d_f$ normalised by the maximum total displacement
$d$, evaluated 5~s after cessation of the laser pulse. \textbf{d)}
Scatter-plot of the radius of gyration $R_g$ (left ordinate, red
symbols); red line shows the average. Right ordinate, black symbols:
maximum number of excited particles; black line shows the average. Laser
power is $99$~mW. \textbf{e)} On log-log scales, the excitation size
$N_\textrm{max}$ is plotted against its radius of gyration $R_g$ where
the shape and color of the data points indicate the laser power. The
phenomenological power-law fit with exponent 1.6 shows that the
excitations of small size, $R_\textrm{g} \leq 6~\mu$m, have a fractal
shape. The box-arrow markers correspond to the excitation patterns shown
in Fig.~\ref{fig:1}. Beyond the power-law, this data shows the abrupt
transition from deterministic response to massive heterogeneity that is
not fractal but just disordered.}
\label{fig:3}
\end{figure}

To characterise this unexpected response in more detail, we determined the
following quantities: the average of $N_{\rm max}$, where $N_\text{max}$
is the maximum of $N(t)$, the standard deviation of $N_{\rm max}$, the
maximum displacement observed in each micro-experiment, $d$, the radial
component of $d$, $d_r$, and the maximum radius of gyration $R_g=[N_{\rm
max}^{-1}\sum_j({\bf r}_j-{\bf r}_{\rm com})^2]^{1/2}$, where ${\bf
r}_j$ and ${\bf r}_{\rm com}$ are the position of particle $j$ and of
the center of mass of the excitation, respectively. In Fig.~\ref{fig:3},
these quantities are plotted against packing fraction. Panel {\bf a)}
demonstrates that $\langle N_{\rm max} \rangle$ is indeed a non-monotonic
function of $\phi$ and that this quantity peaks at around $\phi_{\rm
max}=0.60$. Remarkably, the location of the peak is independent of the
laser power $A$ and the duration of laser (Extended Data Figs.~3{\bf b)},
{\bf c)}, Extended Data Fig.~4), suggesting that this feature represents
an inherent property of the liquid and is not related to the way it is
probed. Note that $\phi_{\rm max}$ is indistinguishable from the onset
packing density $\phi_{\rm onset}=0.6$, i.e.~we have strong evidence that
the non-monotonic behavior of the response occurs at $\phi_{\rm onset}$,
the point at which the cage is formed. At this packing fraction also
$N_{\rm max}$ shows maximum scatter, panel {\bf b)}, indicating that at
$\phi_{\rm max}$ the properties of the excitations depend sensitively
on the local structure of the liquid. At the same packing fraction
the mobile particles move in the most radial manner, panel {\bf c)},
which, in conjunction with Fig.~\ref{fig:1}{\bf f)}, demonstrates that
at $\phi_{\rm max}$ the particle dynamics undergoes a transition from a
viscous response with pronounced radial movement to a solid-like response
with a complex local rearrangement of the particles (see below). Due to
this strong radial component of the motion, the size of the excitation
attains a maximum at $\phi_{\rm max}$, reaching 5-7 particle diameters
and involving the maximum number of particles with both quantities having
a large variance, panel {\bf d)}. A plot of $N_{\rm max}$ vs. $R_g$
shows, panel {\bf e)}, that at small $R_g$ the excitations have a fractal
geometry with a fractal exponent around 1.6 and that this power-law is
an upper bound for $N_{\rm max}$ even for large values of $R_g$. Note
that this behavior is independent of the laser intensity (symbols with
different shape) and hence of the details of the perturbation. For packing
fractions around $\phi_{\rm max}$ the data points are close but slightly
below this power-law, showing that at $\phi_{\rm max}$ the excitation
have a fractal shape and are also very heterogeneous. For larger $\phi$
the data points scatter strongly, i.e.~the excitations no longer are
fractal but are just complicated disordered objects.

\begin{figure}[!h]
\centering
\includegraphics[width=1.0\columnwidth]{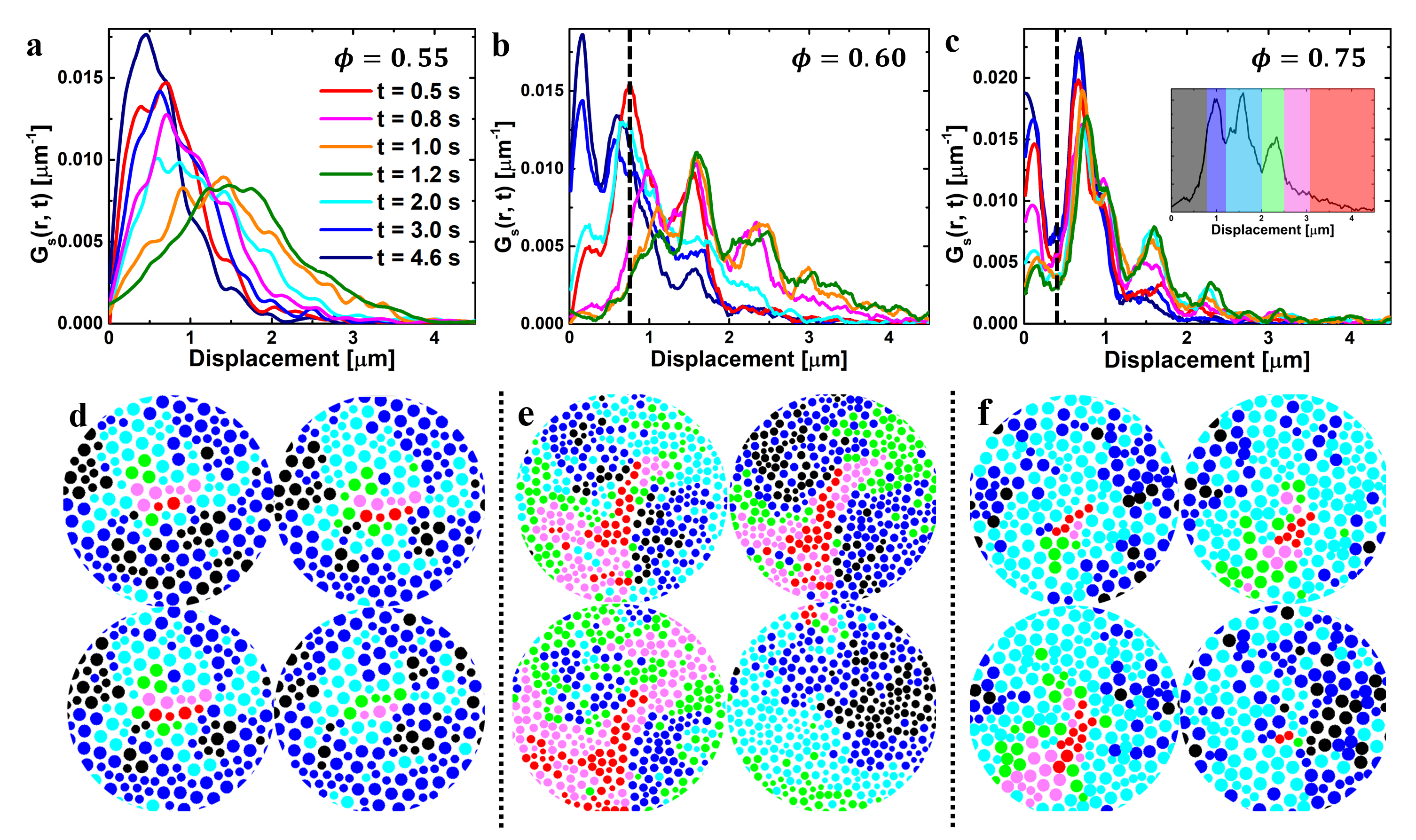}
\caption{\textbf{Figure 3 $|$ Particle displacements for different packing
fractions:} Self part of the van Hove correlation function for different
times (see legend) \textbf{a)} $\phi=0.55$, \textbf{b)} $\phi=0.60$,
\textbf{c)} $\phi=0.75$. The vertical dashed lines in {\bf b)} and {\bf
c)} show the size of the cage as obtained from the MSD. {\bf d)}-{\bf f)}
Displacement field at $t=0.8$~s, $1.0$~s, $1.2$~s, and $2.0$~s (clockwise
from upper left disk). The color gives the displacement with the code
given in the inset of panel {\bf c)}.}
\label{fig:4}
\end{figure}

To quantify the displacement of the mobile particles we use the
self part of the van Hove function~\cite{hansen86}, $G_s(r,t)=n^{-1}
\sum_{j=1}^n \langle \delta(r-|r_j(t)-r_j(0)|)\rangle$, where $\delta$ is
the $\delta-$function, $n$ is the number of particles having a distance
less that 7$\sigma_s$ from the laser spot, and $t=0$~s corresponds to
the time the laser is turned on, Fig.~\ref{fig:4}. For $\phi = 0.55$,
panel {\bf a)}, the particles get displaced significantly, see curve
for $t=1.2$~s, but then return close to their initial position at later
times, thus showing the presence of elastic behavior. We find surprising
that the distributions for $\phi=0.60$, panel {\bf b)}, show a multitude
of regularly spaced peaks indicating that the particles move only by
multiples of a distance $\approx 0.8~\mu$m. Peaks in $G_s(r,t)$ have
been observed previously in the $\alpha-$relaxation regime of deeply
supercooled liquids and have been interpreted as the hopping motion of
{\it individual} particles, which is the relevant transport mechanism in
the deep glassy state~\cite{2011Glassy_Binder,2007_Schweizer}. In these
cases, however, the spacing of the peaks corresponds to the size of a
particle while here it is only about 30~\%, i.e.~the motion occurs on a
significantly smaller scale. Most remarkably, the distributions for long
times, $t \geq 2.0$~s, but still much shorter than the $\alpha-$relaxation
time, are concentrated at distances that are smaller than the ones for
shorter times, $t \approx 1.2$~s, showing that the particles recoil even
after having made large excursions. If $\phi$ is increased to 0.75,
panel {\bf c)}, the peaks at small $r$ become more pronounced while
the ones at large distances decrease, i.e.~the response of the system
signals that the cage has becomes more solid.

To see the cooperative nature of the particle dynamics we show in
panels {\bf d)}-{\bf f)} snapshots of individual excitations at four
different times. Different colors correspond to the location of the
different peaks, see color bar in the inset of panel {\bf c)}. One
recognises that the particles with similar displacements, i.e. same
color, form clusters demonstrating that the motion of the particles is
cooperative and heterogeneous in space. For $\phi=0.60$ these clusters
become very extended at intermediate times, indicating the presence
of zones that are mechanically solid, and hence move as a block,
and coupled only weakly to surrounding zones, thus rationalising the
presence of the many narrowly spaced peaks in $G_s(r,t)$. Although
earlier studies have identified heterogeneous and cooperative motion
in the $\alpha-$relaxation regime at around the packing fraction which
corresponds to the mode-coupling point, i.e.~much higher than $\phi_{\rm
onset}$~\cite{1997PRL_Kob,2000Science_Weeks,2000Science_Kegel}, that
dynamics was irreversible while the motion we document here occurs at
lower packing fraction and is to a large extent reversible, i.e.~the
height of the peak in $G_s(r,t)$ at small $r$ is non-monotonic in
time. This difference reflects the fact that at $\phi_{\rm onset}$
the system is already elastic but still has appreciable local mobility
while  at higher $\phi$ this mobility is strongly suppressed. These data
suggest that phenomena identified earlier with the $\alpha$-relaxation
show an interesting counterpart already at $\phi_{\rm onset}$.

Our micro-rheological experiments show the complexity of cage formation
close to the onset packing fraction. The data suggest that mechanically
stiff regions join to build the dynamical wall surrounding each
particle. In this view the onset packing is the point at which these
(odd shaped) bricks are already quite solid but have not yet merged,
giving rise to large fluctuations in the mechanical response. This
complex mechanism agrees with the theoretical view that cage-formation
is a highly non-linear process involving a feedback mechanism of many
particles~\cite{2008_Gotze}. We expect the observed non-monotonic behavior
in the micro-mechanical response to be general since it is directly
related to the change of the transport mechanism, from a normal fluid-like
behavior for which one has mainly viscous or viscoelastic relaxation
to a dynamics that is more glassy, i.e.~shows a two step relaxation
and has a noticeable elastic component. Although previous studies
showed that glass-forming systems can have a non-monotonic dependence
of quantities like the $\beta-$relaxation time or the size of mobile
regions~\cite{1993PRL_Megen,2012NP_Kob}, these effects occurred around
$\phi_{\rm MCT}$, i.e.~at densities that are significantly higher than the
onset density at which we find a peak in the non-linear response. Hence
these two behaviors should not be confused with each other. Obtaining
a theoretical description of the system's response around the onset
point will be challenging because the local heterogeneities preclude the
luxury of taking a mean-field approach~\cite{2009_Gazuz,2020_Gruber}. One
possibility to advance on this is to connect the local properties of
the potential energy landscape with the relaxation dynamics of the
system~\cite{2018JCP_Lerner} or to use computer simulations to study
the non-linear local response. To link the current micro-experiments
with macroscopic responses, for which so far no non-monotonic behavior
of the observables have been found, will be important since it will
allow to understand what distinguishes a normal liquid from its glassy
counterpart and ultimately why glasses are solid.


\newpage

{\bf Figures with legends:}

\newpage
\clearpage

\textbf{Acknowledgements:} 
We thank L. Cipelletti and M.~D. Ediger for discussions.  This study
was supported by the Korean Institute for Basic Science (project code
IBS-R020-D1) and grant ANR-15-CE30-0003-02. W.~K. is senior member of
the Institut Universitaire de France.

\vspace{12pt}
\textbf{Author contributions:} 
BL, WK, and SG designed the research. BL and KL did the measurements.
BL analyzed the data. BL, WK, and SG wrote the paper. 

\vspace{12pt}
\textbf{Competing interests:} 
The authors declare no competing interests.

\vspace{12pt}
\textbf{Materials \& Correspondence:} 
Correspondence and requests for materials should be addressed
to B.L. (libotc@gmail.com), W.K. (walter.kob@umontpellier.fr) or
S.G. (sgranick@gmail.com).

\newpage

\setcounter{page}{1}

\textbf{\large Methods} 

\textbf{Experimental system}\\[-5mm]

Our glass-former is a binary mixture of anionic PMMA colloidal particles
(Microparticle GmbH) with size $\sigma_\textrm{s}=(2.08 \pm 0.05)~\mu$m
and $\sigma_\textrm{l}=(2.91 \pm 0.05)~\mu$m and number ratio of
$0.55:0.45$. This same system was studied using other approaches in
Ref.~\cite{2018CP_Li}. The particles are slightly negatively charged
and the anionic surface charge comes from sulfate groups brought in
by the initiator in the polymerization reaction. The zeta potential is
around -20~mV. The interaction is hard-sphere like \cite{2019PRX_Li},
which can be recognized by measuring the radial distribution function
of a semi-dilute sample. The Debye length of this system is smaller
than 20~nm as can be estimated from the position of the first peak of
$g(r)$ and the zeta potential~\cite{2019PRX_Li,2017SM_Yang}. Suspended
in water these particles are confined between two glass plates separated
by 3.37~$\mu$m. This spacing is fixed by adding to the suspension a small
amount ($<$0.1$\%$ in number) of 3.37$~\mu$m gold particles (Microparticle
GmbH). It can be expected that the confinement by the two glass walls
will not induce  any additional heterogeneity to the particle dynamics
\cite{2012PRE_Edmond} because the distance between the two glass walls
is uniform. Since the spacing is larger than the size of the colloidal
particles the latter get stuck between the plates only rarely (on average
less than one in a field of view that contains 1000-3000 particles)
and hence the dynamics of the system is not influenced in a significant
manner. As the separation is small, most of the hydrodynamic forces
between the particles are screened and hence we are left with just the
steric hindrance interaction~\cite{2013PRL_Ma}. The roughness of the glass
substrate on the length-scale of microns is on the scale of nanometer
while on the scale of a few hundred~$\mu$m is at most a small fraction of
the size of the gold particles.  To avoid enthalpic interactions between
the particles and the walls we have treated the substrate in the same
way as documented by Refs.~\cite{2018CP_Li,2013PRL_Ma}. Finally we note
that this binary system is a good glass-former and we did not observe
any signs of crystallisation~\cite{2018CP_Li}.

We probe the static and dynamic properties of the system at eight
area fractions $\phi=\frac{1}{4} \pi (\sigma_\textrm{s}^{2}
\rho_\textrm{s}+\sigma_\textrm{l}^{2} \rho_\textrm{l})$ with
$\rho_\textrm{s}$ and $\rho_\textrm{l}$ the number density of small
and large particles, respectively, with $\phi-$value 0.45, 0.52, 0.57,
0.60, 0.65, 0.73, 0.78, and 0.83 (error bars are $\pm$0.01). We use
video microscopy (10~frames/s) to obtain the position of the particles
as a function of time (spatial resolution = 0.1~$\mu$m), using a CMOS
Hamamatsu camera (C11440-42U30). During the 2-to 6-hours measurements
at each $\phi$, no drift, flow, or density change was observed. The
centre-of-mass positions of the colloidal particles is tracked using an
image-processing algorithm \cite{1996ARPC_Murray}.\\[5mm]

\textbf{Probing the response}\\[-5mm]

To probe the response of the system we use focused, pulsed laser
beams (repetition rate 80 MHz) of duration 0.5~s to generate a local
excitation. In practice a spatial light modulator is used to split
one beam into four and we shine these sub-beams simultaneously to
different positions of the sample (avoiding the gold particles).
The distance between these spots is sufficiently large (more than
20~$\sigma_\textrm{s}$) that excitations do not interact, Extended
Data Fig.~1{\bf a)} and Video~1. The time separation between two
excitations is much larger than the relaxation time of the glass-former
at the corresponding $\phi$, i.e.~these events can be considered to be
independent. This setup allows to collect rapidly the necessary large,
statistically-meaningful amounts of data for the subsequent analysis. For
each pair ($\phi,A$), more than 20 individual excitation events were
evaluated to ensure reliable statistics.

The laser beams impinging on the sample will scatter from the colloidal
particles and will exert a dielectrophoretic force to the particles near
the edge of the laser waist, due to the gradient of the electromagnetic
field, pushing them away from the beam center \cite{1999CSAPEA_Peter}. As
a result, the particles very close to the beam start to collide with
their neighboring particles, setting up an excitation.

Previous studies have shown that with this setup of the laser beam
the scattering force is localized to a range that is smaller than two
particle diameters~\cite{2015OE_Villadsen} and to have a magnitude
which is proportional to the laser power $A$ \cite{2015OE_Villadsen,
2016APhot_Zensen}. To get a better understanding on how the laser
influences the motion of a particle we show in Extended Data Fig.~1{\bf
b)} the time dependence of the displacement of a particle that at time
zero is hit by the laser beam. The packing fraction is $\phi=0.35$
and the different curves correspond to particles that at $t=0$ had
different distances from the beam center. (Here the laser power is
$A=99$~mW.) If the initial distance of the particle is about the half
width of the beam the force is very large, since the gradient in the
intensity is at its maximum and the particle gets displaced by around
3.6~$\mu$m. If the initial distance is increased to 0.8~$\sigma_{\rm s}$,
the final displacement is reduced somewhat. In contrast to this we see
that for an initial displacement of 1.18~$\sigma_{\rm s}$ the particle
does not move at all, which shows that the influence of the laser beam
is restricted to a range that is smaller than 1.2~$\sigma_{\rm s}$.

Having demonstrated that the laser influences the motion of a particle
only at small distances one can wonder whether the properties of an
excitation depend on the distance between the beam center and the hit
particle. In Extended Data Figs.~1{\bf c)} and {\bf d)} we thus show
the maximum number of mobile particles, $N_{\rm max}$, and the radius of
gyration, $R_g$, as a function of this distance. One recognizes that there
is no obvious trend in the data, which is evidence that the mentioned
distance does not influence the properties of the resulting excitation.

Extended Data Fig.~1{\bf e)} demonstrates that the maximum intensity
of the laser increases quickly with $A$. However, panel {\bf f)} shows
that the half width of the beam waist remains constant and therefore we
conclude that the region directly influenced by the laser is independent
of its intensity $A$.

Extended Data Fig.~1{\bf g)} presents the long time displacement of
a particle, $D_{\rm max}$, as a function of the laser power used to
hit the particle. The packing fraction is low, $\phi=0.32$, and so
one expects that only the time-dependent dielectrophoresis force from
the laser and the viscous (Stokes) forces act on the particles. (The
former only while the laser is on, i.e.~0.5~s, blue area in Extended Data
Fig.~1{\bf b)}.  We recognize that $D_{\rm max}$ is a linear function in
the laser intensity $A$, a linear fit gives $D_{\rm max}(A)$~[$\mu$m] =
0.036$A$~[mW], showing that $A$ is directly related to the
velocity of the particle at the end of the laser pulse. To estimate
this velocity $v_0$ we use Extended Data Fig.~1{\bf b)} to obtain from
the short time displacement for $d=0.53~\sigma_{\rm s}$ (red curve)
that $v_0=D_{\rm max}/t_{\rm d}$, where $t_{\rm d}=0.5$~s is the
duration of the pulse, i.e.~$v_0\approx 7.2$~$\mu$m/s. Combining this
with the mentioned linear dependence of $D_{\rm max}$ on $A$ we thus
find $v_0~[{\rm \mu m/s}]=0.072\times A~[{\rm mW}]$ which gives for,
say, 100~mW a velocity of 7.2~$\mu$m/s. Using the fact that the work
done by the dielectrophoresis force is the same as the one done by the
viscous force, we estimate the magnitude of the dielectrophoresis force,
averaged over 0.5~s, to be in the pN range.\\[5mm]

{\bf Static and dynamic properties of the quiescent system}\\[-6mm]

In this section we present the static and dynamic properties of our
liquid under quiescent conditions and demonstrate that our system has
the usual characteristic of a standard glass-former.

In Extended Data Fig.~2{\bf a)} we show the radial distribution function
$g(r)$ for the packing fractions considered~\cite{hansen86}. As for all
good glass-formers, the $\phi-$dependence of $g(r)$ is mild in that an
increasing packing fraction mainly results in an increase of the various
peaks~\cite{2011Glassy_Binder}. This is demonstrated in the inset of
the figure which shows the height of the main peak as a function of the
packing fraction. One recognises that this height is a smooth function
of $\phi$, indicating that the system is indeed a good glass-former that
shows no sign of crystallisation.

The simplest way to characterise the relaxation dynamics of a liquid is
via the mean squared displacement (MSD) of the particles

\begin{equation}
\Delta^2 (t) =\frac{1}{N}\sum_{j=1}^N \langle [{\bf r}_j(t) - {\bf r}_j(0)]^2 \rangle.
\label{eq:s1}
\end{equation}

In Extended Data Fig.~2{\bf b)} we show the time dependence of the MSD
for different packing fractions. One recognises that for $\phi=0.45$ the
MSD shows no sign of the presence of a plateau in that it is diffusive
at all times, i.e.~the MSD follows a power-law with exponent 1.0. For
$\phi=0.73$ one finds that at intermediate times (2-100 seconds) the MSD
is almost flat, i.e.~the particles are trapped within their cage, while
at longer times they start to become diffusive. The packing fraction
at which the MSD shows the first significant deviation from a purely
diffusive behavior is around $\phi=0.60$, which is thus a first estimate
for the value of the onset density.

To identify the onset point in a more quantitative manner we take the
first derivative of the MSD. If a particle is caged it will show in the
MSD a sub-diffusive time dependence, i.e.~the local slope will be smaller
than 1.0. In Extended Data Fig.~2{\bf c)} we show this slope and one sees
that for packing fractions up to 0.57 this slope is close to unity for
all times, i.e.~the system is always diffusive. For $\phi=0.60$ a clear
dip occurs and hence we can conclude that this $\phi$ is close to the
onset packing fraction. For higher $\phi$ the dip becomes more pronounced
and extends to longer times, showing that the cage is getting stronger,
as expected for a system that approaches its glass transition.

In the main text we have shown that during an excitation event
the van Hove function of the system shows several peaks that
becomes most pronounced at $\phi_{\rm max}$ and that these peaks
are related to a highly cooperative motion which involves 20
to 50 particles (Fig.~\ref{fig:4}). Extended Data Fig.~2{\bf d)}
demonstrates that in the quiescent system no such peaks are observed,
in agreement with many previous studies of the relaxation dynamics
of colloidal systems and computer simulations of moderately glassy
fluids~\cite{2012RPP_Hunter,1995PRE_Kob}, which is evidence that our
system has no unusual dynamical features.

Another useful observable to probe the dynamics is the self intermediate scattering function~\cite{hansen86},

\begin{equation}
F_s(q,t) =\frac{1}{N}\sum_{j=1}^N \langle \exp[i{\bf q}\cdot ({\bf r}_j(t) - {\bf r}_j(0))] \rangle \quad ,
\label{eq:s2}
\end{equation}

\noindent
where ${\bf q}$ is the wave-vector. Since a liquid is isotropic,
the right hand depends only on the norm $q=|{\bf q}|$. Extended
Data Fig.~2{\bf e)} shows the time dependence of $F_s(q,t)$ for a
wave-vector that corresponds to the maximum in the static structure
factor, i.e.~$q=2.2$~$\mu$m$^{-1}$. We see that for small $\phi$
the time correlation function decays quickly to zero while with
increasing packing fraction the relaxation dynamics quickly slows down,
as expected. For intermediate $\phi$ one notes that the shape of the
time correlation function changes in that it shows at intermediate
times the formation of a shoulder. This feature becomes more pronounced
if $\phi$ is increased more and demonstrates the formation of the
cage that temporarily traps the particles on intermediate time
scales~\cite{2011Glassy_Binder,1995PRE_Kob}. The packing fraction
at which this shoulder starts to become visible is around 0.60,
indicating that this value corresponds roughly to the packing
fraction at the onset point, a result that is compatible with earlier
studies of two-dimensional glass-forming systems~\cite{2010PRE_Kurita,
2017PNAS_Vivek}. From that graph it is also evident that $\phi_{\rm MCT}$,
the critical packing fraction of mode-coupling theory~\cite{2008_Gotze},
is significantly higher than 0.6. Although we have not made a detailed
study to determine $\phi_{\rm MCT}$, previous computer simulations and
experiments suggest that for a two dimensional system $\phi_{\rm MCT}$
is around 0.79~\cite{2011PRE_Weysser,2017PNAS_Vivek}, a value that is
compatible with the dynamic data we have. Below we use the
self intermediate scattering function to obtain the $\alpha$-relaxation
time $\tau_\alpha$. Fitting a Vogel-Fulcher-Tammann-like expression to
$\tau_\alpha(\phi)$, we obtain a $\phi_g$, the packing fraction at the
glass transition, that is around $0.90$.

A further property that characterises glass-forming systems is that
the relaxation dynamics becomes heterogeneous upon  entering the
glassy regime~\cite{2000_Ediger}. This heterogeneity can be seen,
e.g., by probing the non-Gaussian parameter $\alpha_2(t)$ which for a
two-dimensional system is given by

\begin{equation}
\alpha_2(t)= \frac{\langle r^4\rangle}{2\langle r^2\rangle^2}-1 \quad .
\label{eq:s3}
\end{equation}

Extended Data Fig.~2{\bf f)} shows the time dependence of
$\alpha_2$ for the different packing fractions. In agreement
with previous studies one finds that at low $\phi$ the function
is basically zero, i.e.~the system shows no heterogeneity by this
definition~\cite{2000_Ediger,1997PRL_Kob,1995PRE_Kob}. At intermediate
and large $\phi$ one sees a peak in $\alpha_2(t)$, the position of
which moves to larger times with increasing packing fraction and
also its height becomes larger, indicating the increasing presence of
dynamical heterogeneities. This peak starts to become noticeable at
around $\phi=0.60$, indicating that this is indeed close to the onset
packing fraction.

Finally we note that our MSD, $G_s(r,t)$, and $F_s(q,t)$ look
qualitatively similar to the ones found in previous studies of colloidal
systems~\cite{1998_vanmegen,2012RPP_Hunter,2000Science_Weeks,2009_Brambilla}
which is strong evidence that our system shows the typical slowing
down of the relaxation dynamics of a glass-forming system, i.e.~it
can be considered as generic. The dynamic observables indicate that
the onset packing fraction is around 0.60 and also this result
is compatible with previous studies of two dimensional colloidal
systems~\cite{2015PRX_Klix}.\\[5mm]

{\bf Influence of the perturbation on the response of the system}\\[-6mm]

In this section we will discuss how the duration and intensity of the
laser beam influences the response of the system.

In the main text we have shown that the response of the system, as given
by the number of particles that make a significant displacement, shows
a non-monotonic dependence as a function of the packing fraction. These
results were obtained by shining the laser for a period of 0.5~s onto the
sample. Since in complex systems a response can show a maximum because
of the crossing of two time scales (e.g.~the time scale for the driving
and an internal time scale of the system, giving rise to a resonance
phenomenon) it is important to check whether in our case the observed
maximum in the response is indeed an intrinsic property at the packing
fraction $\phi_{\rm max}$ and not just a crossing of time scales. To
elucidate this point we have repeated some of the experiments by using
a time scale for the laser pulse of 1~s, i.e.~twice the normal time. In
Extended Data Fig.~3{\bf a)} we show the relaxation time $\tau_\alpha$
of the system as a function of $\phi$. (Here we use the usual definition
for $\tau_\alpha$ given by $F_s(q,\tau_\alpha)=1/e$.) From this figure
we see that a change in relaxation time by a factor of two corresponds
to a change in the packing fraction by about 0.03, taking $\phi=0.60$
as a reference packing fraction. In Extended Data Fig.~3{\bf b)} we plot
the average number of particles participating in an excitation, $\langle
N_{\rm max}\rangle$, as a function of $\phi$. (Here the laser intensity
is $A=99$~mW and the definition of $N_{\rm max}$ is the same as in the
main text.) One recognises that the curves corresponding to the laser
pulse duration 0.5~s are slightly lower than the one for the duration
1.0~s, a result that makes sense since one injects less power into the
system. More important is, however, the observation that the two curves
show a peak at the same packing fraction and this with a precision that is
definitely better than 0.03. The same conclusion is reached by monitoring
the standard deviation of $N_{\rm max}$, shown in panel {\bf c)}. Hence
we can conclude that the packing fraction at which one finds a maximum
response is not influenced in a significant manner by the duration of
the laser pulse, i.e.~$\phi_{\rm max}$ is a property of the system and
not of the manner by which it is determined. From panels {\bf b)} and
{\bf c)} one also recognizes that the increase of $N_{\rm max}$ and its
variance is faster than linear for $\phi < \phi_{\rm max}$. This is thus
evidence that the growth of these quantities reflects a response of the
system related to an increasing collective behavior of the particles if
$\phi$ is increased and is not just due to the fact that at higher $\phi$
the laser beam hits on average more particles.

A second important parameter for the local perturbation is the intensity
of the laser beam. Extended Data Fig.~4{\bf a)} shows, for different
packing fraction, the average size of the excitation, $\langle N_{\rm
max} \rangle$, as a function of the laser pulse intensity $A$. We see
that for all values $A$ considered, $\langle N_{\rm max} \rangle$ shows
a  maximum at around $\phi_{\rm max}=0.60$. This demonstrates that the
packing fraction at which the system has a maximum response is independent
of the laser intensity. The same conclusion is reached by looking at the
$A-$dependence of the variance of $N_{\rm max}$, Extended Data Fig.~4{\bf
b)}. Thus we conclude that the maximum response of the system to an
external perturbation does not depend in a significant manner on the
details of the perturbation. Note that the $A-$dependence of $\langle
N_{\rm max} \rangle$ is linear. However, a linear fit to the data point
intercepts at $A=0$ the ordinate at a negative value of $\langle N_{\rm
max} \rangle$. Since the fits cross the abscissa at around 10~mW, a value
that is smaller than the ones we use in our experiments, we have evidence
that we are not probing the linear response of the system, but are in
the non-linear regime, in agreement with the observation that for all
values of $A$ we use we do see plastic events. From this graph it can
be concluded that for laser intensities less than 10~mW the reaction of
the system is elastic, i.e.~no excitation will be generated.

In Extended Data Fig.~4{\bf c)} we present a three-dimensional plot that
collects the information on how the size of the excitation depends on
the laser intensity and the packing fraction. In agreement with the
data shown in panel {\bf b)} we find that if $A$ is kept constant
$\langle N_{\rm max}\rangle$ shows a maximum at a packing fraction
that is around 0.60, i.e.~it is independent of $A$. Also included in
the graph is the information regarding the variance of $N_{\rm max}$
(coded via the color). One sees that a constant $N_{\rm max}$ implies
the same color, i.e.~the variance is constant as well. So one concludes
that the size of the excitation determines its variance, i.e.~the latter
depends on $\phi$ and $A$ only via $N_{\rm max}$.

Finally we show in Extended Data Fig.~5{\bf a)} a snapshot of an
excitation produced with a beam of low intensity, $A=25$~mW, in a sample
with intermediate density, $\phi=0.58$. After a waiting time of around
20~s, i.e.~a time after which the excitation has completely died away,
a second pulse is fired at the same spot in the sample. One finds,
panel {\bf b)}, that the pattern of the excitation is very similar to
the one of the first excitation, which shows that the overall shape of
the excitation is encoded in the local structure of the liquid. Panel
{\bf c)} shows the result of a third laser pulse at the same spot
and one finds again that the shape of the excitation does not change,
thus confirming that the local structure of the liquid determines the
displacement of the particles after the pulse.

If the intensity of the pulse is increased to $83$~mW, the first pulse
gives rise to an excitation that is larger than the one obtained with
the weaker intensity (panel {\bf d)}, a result that makes sense since the
injected energy is larger and is coherent with Extended Data Fig.~4{\bf
a)}. In this case the pulse perturbs the local structure so strongly
that a second laser pulse hitting 19~s later the same spot will generate
an excitation pattern which is very different from the first one, panel
{\bf e)}, and the same is true if one fires a third laser pulse, panel
{\bf f)}. Since Extended Data Fig.~4 demonstrates that the qualitative
response of the system is independent of $A$ one thus can conclude that
the presence of the irreversible motion induced by a high laser intensity
does not perturb the local equilibrium of the system, or at least not
in a significant manner, i.e.~the laser pulse can indeed be used as
a micro-rheological perturbation that allows to probe the equilibrium
properties of the system.\\[5mm]

{\bf Videos}\\[-6mm]

Video 1: Main idea of the experiment. A femtosecond laser is focused
to the spot marked by a red cross for an interval of 0.5~s starting at
$t=1$~s into the video. This causes the particle near the focal point
to shift position abruptly and we analyze the resulting deformations of
the nearby particles. In this example, the area fraction is 0.35 and
laser intensity at the focus is 99 mW. At this low area fraction, the
particle displacement is rapid and permanent, and neighbors experience
little perturbation.\\[4mm]

Video 2: Data and analysis illustrated below the onset density. In
this example the area fraction is 0.50 and the laser power 99~mW. The
beam of the femtosecond laser is split into 4 focus spots, marked here
by red crosses, for an interval of 0.5~s starting at $t=1$~s into the
video. The focus spots are separated by a distance sufficiently large so
that responses to the perturbation are independent. Right panel:  video
showing raw data. Left panel: associated displacements of neighboring
particles using the color code defined in Fig.~1.\\[4mm]

Video 3:  Data and analysis illustrated at the onset density. In this
example the area fraction is 0.60 and the laser power 99~mW. The beam
of the femtosecond laser is split into 4 focus spots, marked here
by red crosses, for an interval of 0.5~s starting at $t=1$~s into the
video. The focus spots are separated by a distance sufficiently large so
that responses to the perturbation are independent. Right panel:  video
showing raw data. Left panel: associated displacements of neighboring
particles using the color code defined in Fig.~1.\\[4mm]

Video 4:  Data and analysis illustrated above the onset density. In
this example the area fraction is 0.70 and the laser power 99~mW. The
beam of the femtosecond laser is split into 4 focus spots, marked here
by red crosses, for an interval of 0.5~s starting at $t=1$~s into the
video. The focus spots are separated by a distance sufficiently large so
that responses to the perturbation are independent. Right panel:  video
showing raw data. Left panel: associated displacements of neighboring
particles using the color code defined in Fig.~1.\\[4mm]

{\bf Data availability statement:}\\
The figures and videos that support the findings of this
study are available in zenodo with the identifier
https://zenodo.org/record/3989982\#.Xzv\_cxFS8nQ,
doi:10.5281/zenodo.3989982.

{\bf Code availability statement:}\\
The code used in this study is available from B.L.

\newpage

\renewcommand{\thefigure}{S1}
\begin{figure}[ht]
\centering
\includegraphics[width=1.0\columnwidth]{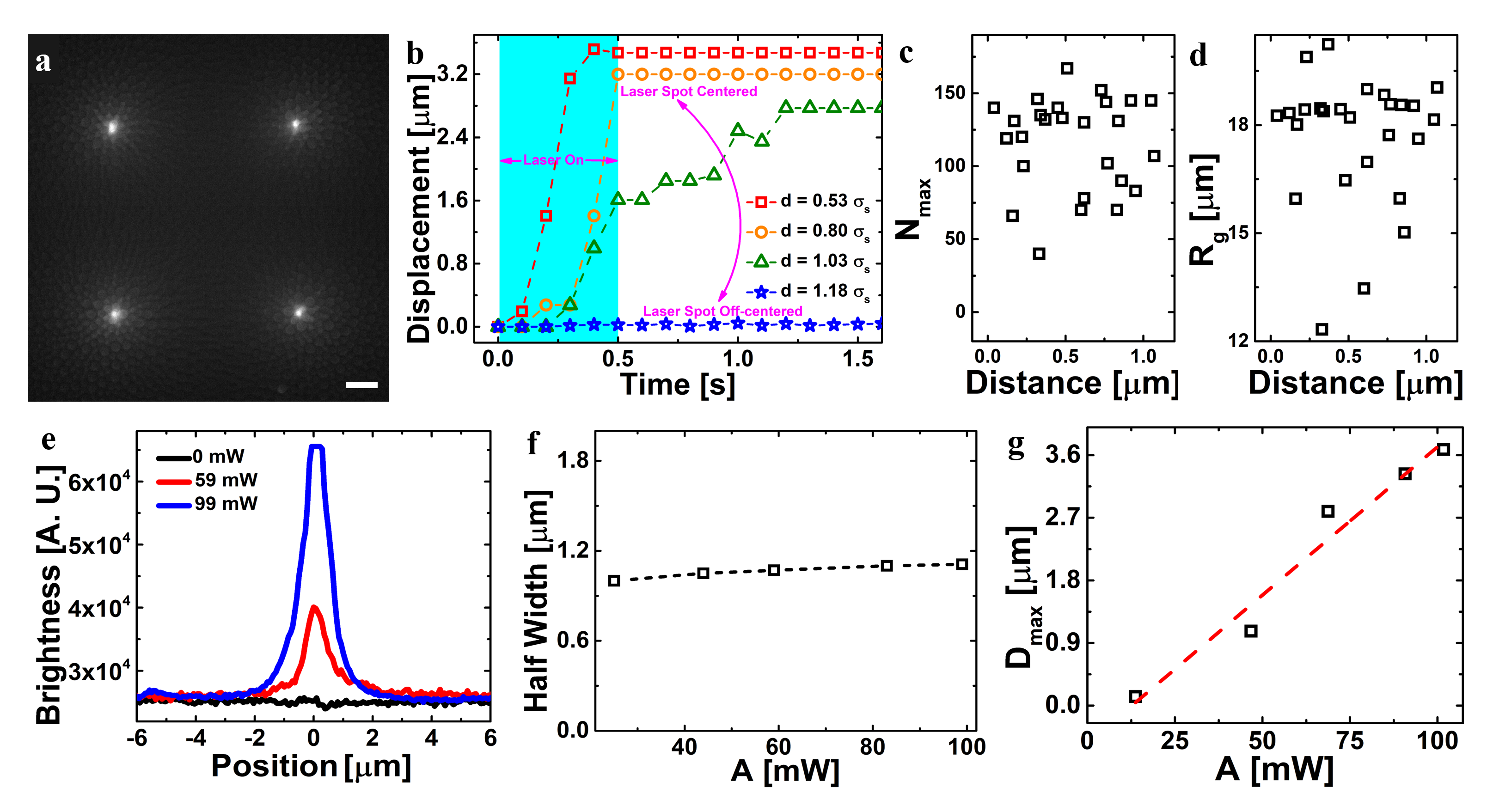}
\caption{\textbf{Extended Data Figure 1 $|$ Details of laser beam setup.}
\textbf{a)} The bright field image of a sample under four-point laser beam
designed to allow parallel accumulation of large amounts of statistics in
the same experiment. The illuminated spots have a distance from each other
that is sufficiently large that the excitations are independent. Scale
bar is 12~$\mu$m. \textbf{b)} The displacement as a function of time
for four typical particles which have different distances to the center
of the laser beam. $\phi=0.35$. To show the direct effect of the laser,
we chose the particles which do not collide with other particles during
the whole excitation event. \textbf{c)} The maximum number of particles
in an excitation as a function of distance between the beam center
and mass center of the particle which is closest to the laser spot,
for $\phi=0.60$ and $A=99$~mW. \textbf{d)} The radius of gyration as
a function of distance between the beam center and mass center of the
particle who is closest to the laser spot, for $\phi=0.60$ and $A=99$~mW.
\textbf{e)} The brightness intensity profile across the laser spot for
two value of the beam power (see legend). The curve for $A=0$ shows
the noise of the signal in the absence of a beam.  \textbf{f)} The
half width of the intensity profile as a function of $A$.  \textbf{g)}
Displacement of an isolated particle which has been hit by a laser pulse
with intensity $A$ and duration 0.5~s. Packing fraction is $\phi=0.32$.
}
\label{fig:S1}
\end{figure}

\renewcommand{\thefigure}{S2}
\begin{figure}[ht]
\centering
\includegraphics[width=0.8\columnwidth]{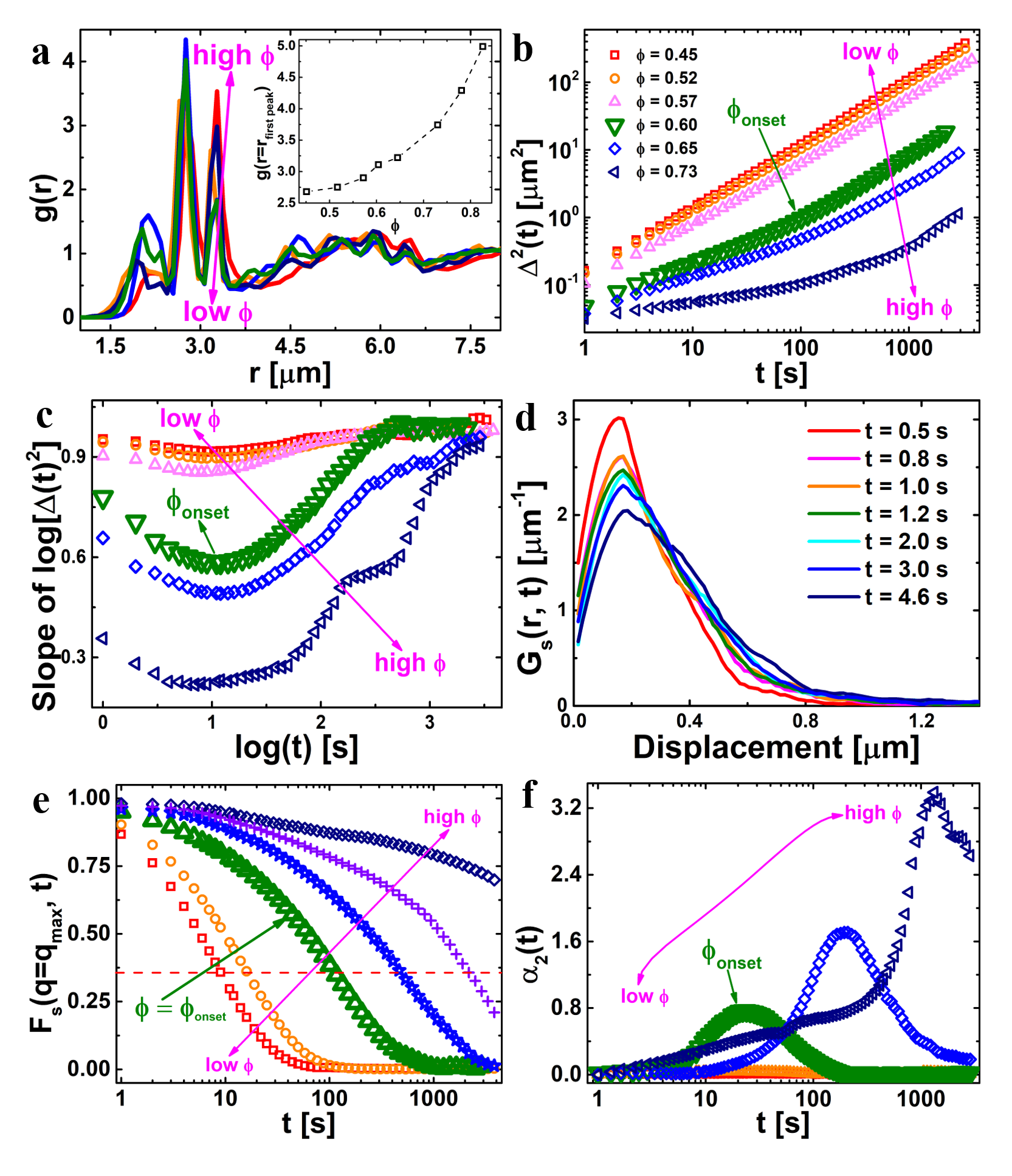}
\caption{\textbf{Extended Data Figure 2 $|$ Static and dynamic
properties of the quiescent system.} $\phi=0.45$, 0.52, 0.57, 0.60,
0.65, and 0.73. \textbf{a)} The radial distribution function $g(r)$ for
samples with different $\phi$. Inset: Height of the first peak of $g(r)$
as a function of $\phi$. \textbf{b)} The mean squared displacement as
a function of time for samples with different $\phi$. \textbf{c)} The
first derivative of the mean square displacement as a function of time
for samples with different $\phi$.  \textbf{d)} The van Hove function
for $\phi=0.60$ at different times (see also Videos 2-4). \textbf{e)}
Intermediate scattering function for samples with different $\phi$. The
wave-vector is $q=2.2~\mu$m$^{-1}$. \textbf{f)} The non-Gaussian parameter
as a function of time for samples with different $\phi$.}
\label{fig:S2}
\end{figure}

\renewcommand{\thefigure}{S3}
\begin{figure}[ht]
\centering
\includegraphics[width=1.0\columnwidth]{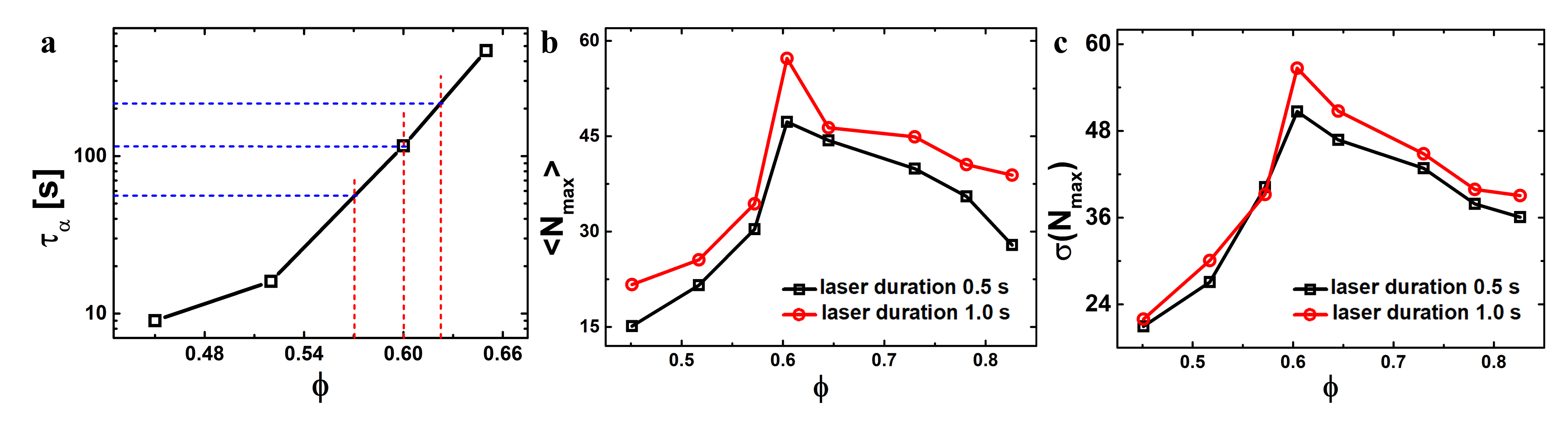}
\caption{\textbf{Extended Data Figure 3 $|$ Influence of laser pulse
duration on excitation pattern.} \textbf{a)} A change of the relaxation
time $\tau_\alpha$, defined as the time at which the intermediate
scattering function in Extended Data Fig.~2{\bf e)} decays to $1/e$,
in log-linear scale, by a factor of two corresponds to a significant
change in the corresponding packing fractions $\phi$. \textbf{b)} Size
of an excitation as a function of $\phi$ for two different values of
the laser duration. For both cases $A=99$~mW. \textbf{c)} Variance of
the maximum number of particles participating in an excitations as a
function of $\phi$.}
\label{fig:S3}
\end{figure}

\renewcommand{\thefigure}{S4}
\begin{figure}[ht]
\centering
\includegraphics[width=0.9\columnwidth]{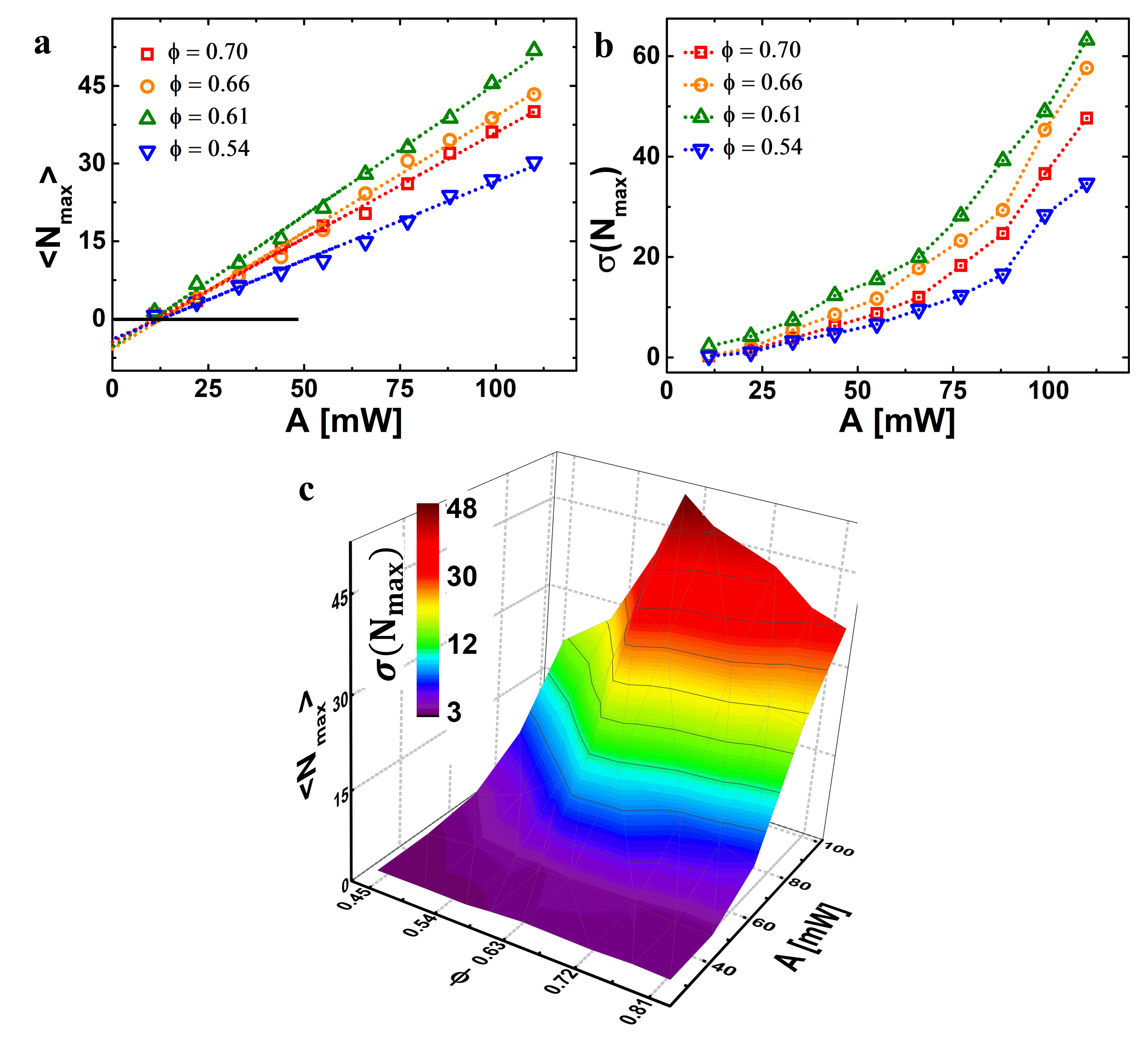}
\caption{\textbf{Extended Data Figure 4 $|$ The non-monotonic response
of the system is independent of the laser intensity.} \textbf{a)}
The averaged excitation size as a function of $A$. \textbf{b)} The
averaged variance of excitation size as a function of $A$. \textbf{c)}
The 3D-plot of excitation size as a function of both $\phi$ and $A$. The
color represents the variance of the excitation size as identified in
the color map.}
\label{fig:S4}
\end{figure}

\renewcommand{\thefigure}{S5}
\begin{figure}[ht]
\centering
\includegraphics[width=1.0\columnwidth]{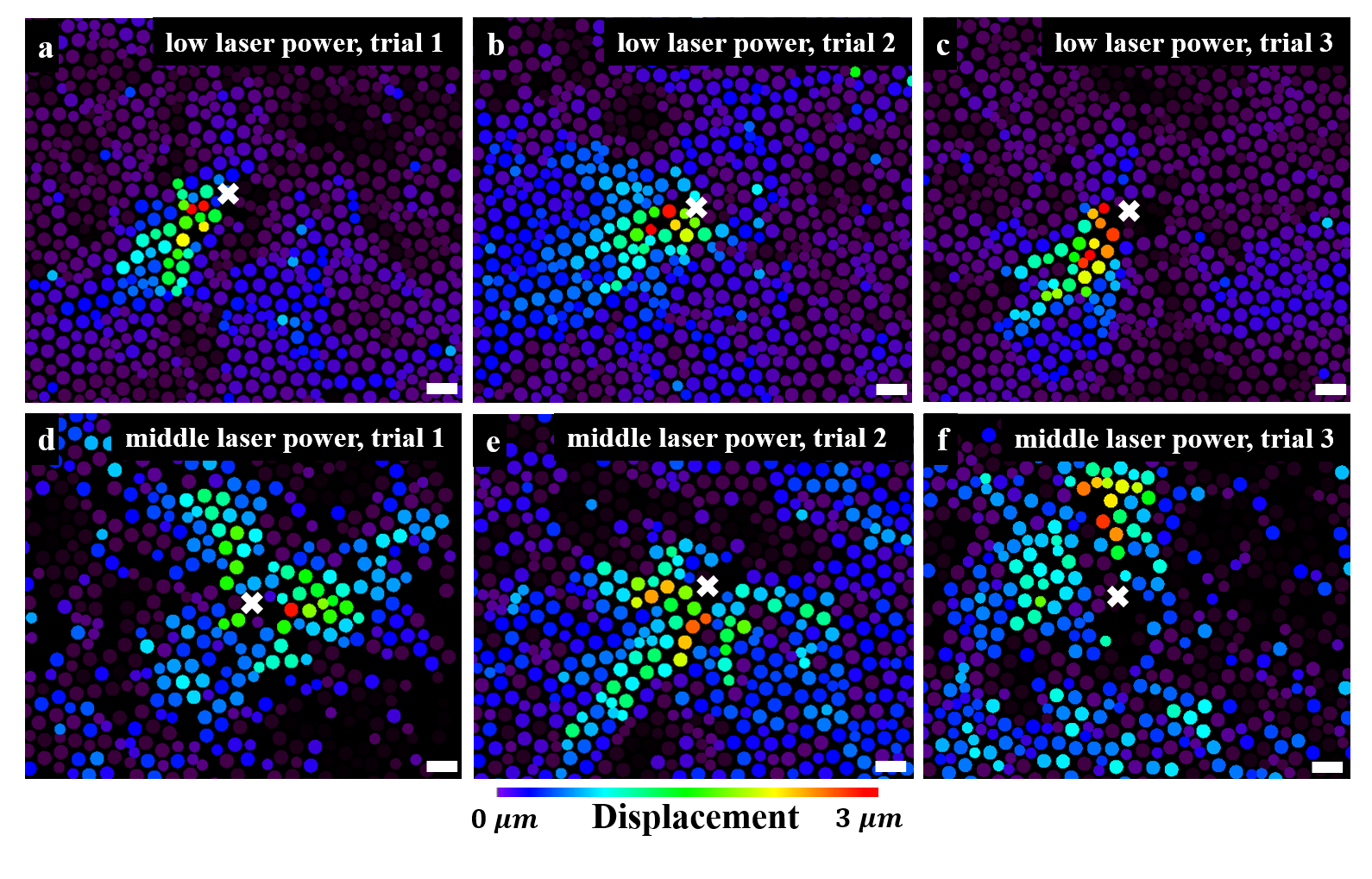}
\caption{\textbf{Extended Data Figure 5 $|$ Excitation patterns for
different laser intensities.} \textbf{a)-c)}Excitation patterns
for A=25~mW in a $\phi=0.58$ sample 5~s after the laser has been
turned off. The color represents the particles' displacements (see
color bar). The duration of the pulse was 0.5~s. The separation
between two excitation events was around 20~s. The scale bar is
10~$\mu$m. \textbf{d)-f)}: Same for A=59~mW.}
\label{fig:S5}
\end{figure}

\clearpage

\end{document}